\documentclass[twocolumn,showpacs,preprintnumbers]{revtex4}
\usepackage{amsfonts}
\usepackage{amssymb}
\usepackage{amsmath}
\usepackage{epsfig}
\usepackage{dcolumn}
\usepackage{bm}

\begin{document}

\date{\today }
\title{\textbf{Bose gas in disordered, finite-layered systems }}
\author{ V.E. Barrag\'an$^{1,2}$, M. Fortes$^{1}$, M.A. Sol\'is$^{1}$, P. Salas$^{1}$}
\affiliation {$^{1}$Instituto de F\'isica, Universidad Nacional Aut\'onoma de M\'exico,
Apartado Postal 20-364, M\'{e}xico D.F., M\'{e}xico}
\affiliation {$^{2}$Posgrado en Ciencia e Ingenier\'ia de Materiales, Universidad Nacional Aut\'onoma de M\'exico, M\'exico D.F., M\'exico}


\begin{abstract}
Disorder effects in the thermodynamic properties of a ideal
Bose gas  
confined in a semi-infinite multi-layer structure 
within a box of thickness $L$ and infinite lateral extent, are analyzed. 
The layers are first modeled by a periodic array of $M$ Dirac delta-functions of equal intensity. Then, we introduce structural and compositional disorder, as well as a random set of layer vacancies in the system to calculate the internal energy, chemical potential and the
specific heat for different configurations. Whereas structural and compositional disorder does not reveal a significant change, a dramatic
increase in the maximum of the specific heat is observed when the system is depleted a fraction of the order of $0.1$ to $0.2$ of random layers compared to the original, fully periodic array. Furthermore, this maximum, which is reminiscent of a Bose-Einstein condensation for an infinite array, occurs at higher temperatures.
\end{abstract}

\maketitle

\address{$^1$Posgrado en Ciencia e Ingenier\'{\i}a de Materiales, UNAM, M\'exico \\
Apdo. Postal 70-360, 04510 M\'exico, D.F., MEXICO\\
$^2$Instituto de F\'{\i}sica, UNAM, Apdo. Postal 20-364,
01000 M\'exico, D.F., MEXICO
}

\section{Introduction}

The thermodynamic properties of a non-interacting Bose gas in layered structures have been studied \cite{SalasJLTP2010,SalasPRA2010,Pathriax,Otros} to understand how the Bose condensate transition is modified by a periodic array of planes in one, two and three dimensions. Layered systems provide a simple model for high-temperature superconductivity (HTSC) materials where the conduction is believed to occur in the well-defined copper-oxide planes. Cooper pairs may be represented by the boson gas in underdoped cuprates where their coherence length is similar to particle separation. When Cooper pairs are allowed to tunnel through permeable layers, the delocalization process of the pairs gives rise to a substantial enhancement of pairing \cite{Chakravarty}. By allowing these systems to be finite in one or two dimensions, it is interesting to explore how the thermodynamic and conduction properties are modified, for example due to the presence of surface states \cite{surfacestates} or through lattice disorder. In the case of a fermion gas (electrons) in a large periodic array, lattice disorder enhances localized states as was shown by Anderson almost sixty year ago \cite{Anderson}. Also, localized states in a Bose-Einstein condensate (BEC) have recently been observed in the context of bichromatic optical lattices obtained by superimposing two one-dimensional optical lattices with different wavelengths, or in waveguides in the presence of a controlled disorder created by laser speckle \cite{Roati,Modugno,Aspect,Aspect2}.

Disorder or random impurities have also been shown to produce BEC in low-dimensional systems in the thermodynamic limit \cite{Luttinger1,Luttinger2} which would otherwise be impossible in an ordered lattice due to the Hohenberg-Mermin-Wagner theorem \cite{H-M-W}. It is therefore interesting to study the properties of a Bose system within a disordered layer structure of finite extent and to understand how the possible formation of a BEC can be enhanced by manipulating the positions of the layers.




In this work, we analyze the wall-disorder effects on the specific heat of an ideal Bose gas  
confined to a semi-infinite layered system described by $M$ permeable barriers within a box of thickness
$L$ and infinite lateral extent. Disorder is created 
when the reference, periodic-layered system goes
to configurations with random barrier positions (\textit{structural} disorder), random barrier permeabilities (\textit{compositional} disorder), or by removing a fraction of the 
walls
at arbitrary sites. 
%
%
%
In the case of wall vacancies, \textit{i.e.}, when the periodicity is broken by the removal of a fraction of the walls,
the effects on the thermodynamic properties of the gas are remarkable. 

Our starting scheme is the derivation of the dispersion relation for a system of $N$ Bose particles in a layered sample where the layers or walls are represented by Dirac-delta functions at fixed or at random distances. 
In Section II we describe our system and derive the dispersion relation using the transfer-matrix method to obtain the energy levels of arbitrary configurations of delta-potential barriers. In Section III, we derive analytic expressions for the internal energy, the chemical potential and for the specific heat of the system. Section IV is devoted to the analysis of plane suppressions or vacancies. Our conclusions are summarized in Section V.

\section{Finite layered system}

We consider an ideal Bose gas confined to a semi-infinite box of thickness
$L$ and infinite lateral extent, divided
into $M+1$ smaller boxes of width $\Delta _{j}=z_{j}-z_{j-1}$ with $%
j=1,..., M+1$ such that the impenetrable border walls are at $z_{0}=0$ and $%
z_{M+1}=L$ where we assume that the wave function vanishes.
\begin{figure}[tbh]
\centerline{\epsfig{file=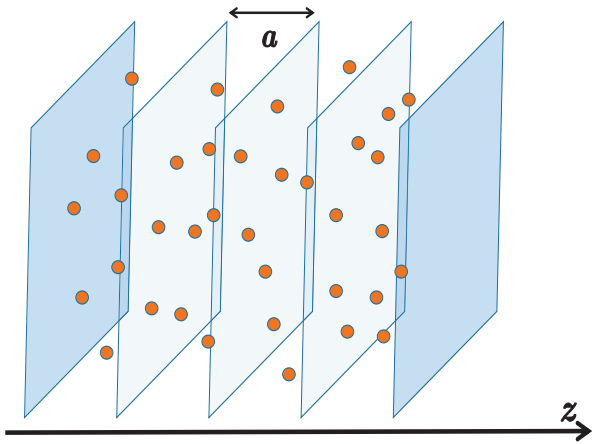,height=3.20in,width=2.5in,angle=-90}}
\caption{(Color online) The ordered, layered-system with permeable planes at fixed
separation $a$. The length of the system in the $z$-direction is $L=(M+1)a$
and infinite in the other two directions.}
\label{fig:Planes4}
\end{figure}
%
Walls or barriers between adjacent boxes are modeled by Dirac-delta
potentials of variable strength in the $z$-direction such that particles are able to 
tunnel through adjacent boxes while they are free
to move in the $x$- and $y$-directions (Fig. \ref{fig:Planes4}). 
The (internal) wall permeability is inversely proportional to the strength of the delta potential .
 

The $M$ Dirac-delta potentials
 in the $z$-direction are given by
\begin{equation*}
V(z)=\sum_{j=1}^{M}v_{j}\delta (z-z_{j}).
\end{equation*}
which in 3D become surfaces of strength $v_j$ located at $z_j$.
%
%
\begin{figure}[tbh]
\centerline{
\epsfig{file=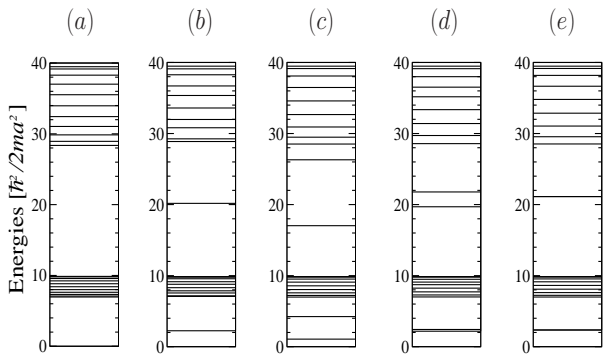,height=1.8in,width=3.1in,angle=0}}
\caption{Energy levels of: (a) ordered system and (b-e) removal of \{1\}, \{1, 2\},
 \{1, 3\}, and \{1, 10\} planes.}
\label{fig:levels}
\end{figure}
For the ordered or periodic, reference system, $\Delta _{j}=a$ and $v_{j}=v$ for all $j.$ Then the dispersion relation in the $z-$direction is given by \cite{Griffiths}
\begin{equation}
\Lambda \frac{\sin (\kappa a)}{\kappa }+\cos \left( \kappa a\right) =\cos
\left( \frac{(i+n-1)\pi }{M+1}\right)  \label{eq:KPreldisp}
\end{equation}%
with $i = 1, 2, 3, \ldots, M+1$, $n=1,2, \ldots$ is the energy band, $\Lambda ={mv}/{\hbar ^{2}}$ and the $z-$direction energy levels are $\varepsilon_{\kappa_{i,n}}={\hbar ^{2}\kappa _{i,n}^{2}}/{2m}$, where $%
\kappa _{i,n}$ are the solutions to Eq. (\ref{eq:KPreldisp}). The band
structure defined by Eq. (\ref{eq:KPreldisp}) is already discernible when the number of planes is as small as $M\sim 10$, as shown in 
Fig. \ref{fig:levels}-($a$). Disorder in the multi-layered system is introduced by allowing the separations $\Delta _{j}$ and/or the strengths $\Lambda _{j}$ to be
different by a random variation of its reference values within a ``noise" interval. In this case, it is more convenient to use the transfer-matrix method relating the wave-functions in the $z-$direction between adjacent regions separated by Dirac delta functions. In the interval $j$ we have $%
\psi _{j}(z)=A_{j}\exp (\mathrm{i}\kappa z)+B_{j}\exp (-{\mathrm{i}}\kappa
z) $ and%
\begin{equation*}
\left(
\begin{array}{c}
A_{j} \\
B_{j}%
\end{array}%
\right) =\mathbb{T}_{j}\left(
\begin{array}{c}
A_{j+1} \\
B_{j+1}%
\end{array}%
\right) ,
\end{equation*}%
where%
\begin{equation}
\mathbb{T}_{j}\mathbb{=}\left(
\begin{array}{cc}
1+\frac{\mathrm{i}\Lambda _{j}}{\kappa } & \frac{\mathrm{i}e^{-2\mathrm{i}%
z_{j}\kappa }}{\kappa } \\
&  \\
-\frac{\mathrm{i}e^{2\mathrm{i}z_{j}\kappa }}{\kappa } & 1-\frac{\mathrm{i}%
\Lambda _{j}}{\kappa }%
\end{array}%
\right) , \hspace{0.2cm} j=1,2,...,M \label{eq:transmat}
\end{equation}%
and we assume box boundary conditions%
\begin{equation}
\psi _{1}(0)=\psi _{M+1}(L)=0.  \label{eq:bc}
\end{equation}%
For a system of $M$ potentials we have%
\begin{equation*}
\left(
\begin{array}{c}
A_{1} \\
B_{1}%
\end{array}%
\right) =\left( \prod\limits_{j=1}^{M}\mathbb{T}_{j}\right) \left(
\begin{array}{c}
A_{M+1} \\
B_{M+1}%
\end{array}%
\right) =\boldsymbol{\tau }\left(
\begin{array}{c}
A_{M+1} \\
B_{M+1}%
\end{array}%
\right)
\end{equation*}%
where $\boldsymbol{\tau }=\prod\limits_{j=1}^{M}\mathbb{T}_{j}$ is a $%
2\times 2$ matrix and the dispersion relation that defines the $z-$direction
energy levels is given in terms of the elements of the product matrix $%
\boldsymbol{\tau }$\ by%
\begin{equation}
\mbox{Im}\left[ \left( \boldsymbol{\tau }_{11}+\boldsymbol{\tau }%
_{12}\right) e^{-\mathrm{i}\kappa L}\right] =0.  \label{eq:disprelT}
\end{equation}%
For a periodic system, Eqs. (\ref{eq:KPreldisp}) and (\ref{eq:disprelT}) are
equivalent. 
In each band, there are $M$ discrete levels due to the barriers
and an additional level at the top of the $n$-band with $\kappa
_{M+1}a=n\pi .$
When a finite number of layers is suppressed an equal number of levels in each band is moved down to the forbidden region. For example, in Fig. \ref{fig:levels} we show the first two energy bands with $M+1=11$ levels when one layer ($b$) or different sets of two layers are removed ($c-e$). Level degeneracy is observed when the removed layers are at symmetrical positions as shown in the diagram of Fig. \ref{fig:levels}-($e$).

\section{Thermodynamic properties}

When there is no interaction between the particles, the partition function
for this system is%
\begin{equation}
\mathbb{Z=}\prod\limits_{n=1}^{\infty}\prod\limits_{i=1}^{M+1}\left( 1-ze^{-\beta \varepsilon
_{i,n}}\right) ^{-1},  \notag  \label{eq:partfcn}
\end{equation}%
where $z=e^{\beta\mu}$ is the fugacity and the single-particle energies are%
\begin{equation*}
\varepsilon _{i,n}=\frac{\hbar ^{2}(k_{x}^{2}+k_{y}^{2})}{2m}+\varepsilon
_{\kappa _{i,n}}
\end{equation*}%
The thermodynamic properties of the system are obtained from the grand
potential
\begin{equation*}
\Omega (T,V,\mu )=-k_{B}T\ln \mathbb{Z=}k_{B}T\sum_{n,i}\ln \left[ 1-e^{-\beta
(\varepsilon _{i,n}-\mu )}\right] .
\end{equation*}

When the system size in the $x$- and $y$-directions is infinite, sums become
integrals and the number equation $N={\partial \Omega }/{\partial \mu }$
becomes
\begin{equation}
N=\sum\limits_{n,i}\left( \frac{L}{2\pi }\right) ^{2}\int
dk_{x}dk_{y}\left\{ \alpha _{\kappa _{i,n}}^{-1}\exp \left[ \frac{\lambda ^{2}%
}{2\pi }(k_{x}^{2}+k_{y}^{2})\right] -1\right\} ^{-1}  \label{eq:numeq}
\end{equation}

where%
\begin{eqnarray*}
\alpha _{\kappa _{i,n}} &=&e^{\beta \left( \mu -\varepsilon _{\kappa
_{i,n}}\right) }, \\
\lambda ^{2} &=&\frac{h^{2}}{2\pi mk_{B}T}.
\end{eqnarray*}%
Integration over $k_{x},$ $k_{y}$ yields
\begin{equation*}
N=-\left( \frac{L}{\lambda }\right) ^{2}\sum\limits_{n=1}^{\infty}\sum\limits_{i=1}^{M+1}\ln
(1-\alpha _{\kappa _{i,n}}),
\end{equation*}%
which may be expressed in terms of the ideal boson gas (IBG) condensation
temperature $T_{0}$ and the corresponding thermal wavelength $\lambda _{0}=h/\sqrt{2\pi
mk_{B}T_{0}}$ as
\begin{equation}
1=-\frac{\lambda _{0}T}{\zeta (3/2)(M+1)aT_{0}}\sum_{n=1}^{\infty}\sum_{i=1}^{M+1}\ln \left[
1-e^{\beta (\mu -\varepsilon _{\kappa _{i,n}})}\right] .  \label{eq:numeq1}
\end{equation}%
The internal energy $U$ is obtained from%
\begin{equation*}
U=\sum\limits_{n,i}\left( \frac{L}{2\pi }\right) ^{2}\int dk_{x}dk_{y}\frac{%
\varepsilon _{k_{x}}+\varepsilon _{k_{y}}+\varepsilon _{\kappa _{i,n}}}{%
e^{\beta \left( \varepsilon _{k_{x}}+\varepsilon _{k_{y}}+\varepsilon
_{\kappa _{i,n}}-\mu \right) }-1}.
\end{equation*}%
Again, integration over $k_{x},$ $k_{y}$ and using the IBG condensation
temperature yields
\begin{eqnarray}
\frac{U}{Nk_{B}T}&=&\frac{\lambda _{0}}{\zeta (3/2)(M+1)a}\sum_{n,i}\left[ \frac{T}{T_{0}}g_{2}(\alpha _{\kappa _{i,n}})+ \right.  \notag \\
&& \qquad \qquad \qquad \left. \frac{\varepsilon _{\kappa _{i,n}}}{k_{B}T_{0}}%
g_{1}(\alpha _{\kappa _{i,n}})\right] ,  \label{eq:U}
\end{eqnarray}%
where $g_{\sigma}(z)$ is the $\sigma$-th order Bose function \cite{Pathria}.

{The specific heat at constant volume is $C_{V}=\left({\partial U}/{\partial
T}\right) _{V}$ yielding
\begin{eqnarray}
\frac{C_{V}}{Nk_{B}}=\frac{\sqrt{4\pi \gamma }}{\zeta (3/2)(M+1)}%
\sum_{\kappa _{i,n}}\left[ \frac{2T}{T_{0}}g_{2}(\alpha
_{\kappa_{i,n}})+ \right. &&  \notag \\
\left. 2\gamma \overline{\varepsilon }g_{1}(\alpha _{\kappa _{i,n}}) +(\gamma
\overline{\varepsilon }_{\kappa _{i,n}})^{2}\frac{T_{0}}{T}g_{0}(\alpha
_{\kappa _{i,n}})+f(\overline{\mu })\right] \quad && ,  \label{eq:Cv}
\end{eqnarray}%
where $\gamma =\lambda _{0}^{2}/4\pi a^{2}$, 
 $\zeta $ is the Riemann
Zeta-function and
\begin{equation}
f(\overline{\mu })=\left( T\frac{\partial \overline{\mu }}{\partial T}-%
\overline{\mu }\right) \left[ \sqrt{\frac{\gamma }{4\pi }}\frac{T_{0}}{T}%
+\gamma ^{2}\overline{\varepsilon }_{\kappa _{i,n}}g_{0}(\alpha _{\kappa _{i,n}})%
\right] .  \label{eq:fmu}
\end{equation}%
where we have used dimensionless units, namely $\overline{\mu }=\mu /(\hbar
^{2}/2ma^{2})$ and $\overline{\varepsilon }_{\kappa _{i,n}}=\varepsilon
_{\kappa _{i,n}}/(\hbar ^{2}/2ma^{2})$. }

\begin{figure}[tbh]
\centerline{
\epsfig{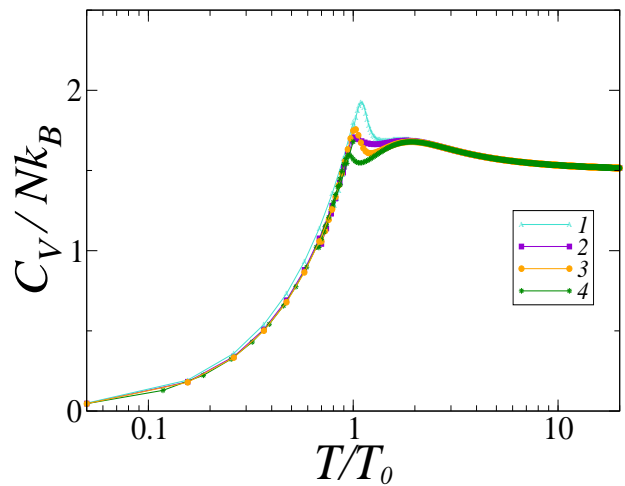}}
\caption{(Color online) Specific heat for random 10 percent variations in the position of the barriers for a system of $M=10$ planes of equal strength, $\Lambda=10$. Each curve denotes a different set of random trials}
\label{fig:Cvstructural}
\end{figure}

\begin{figure}[tbh]
\centerline{
\epsfig{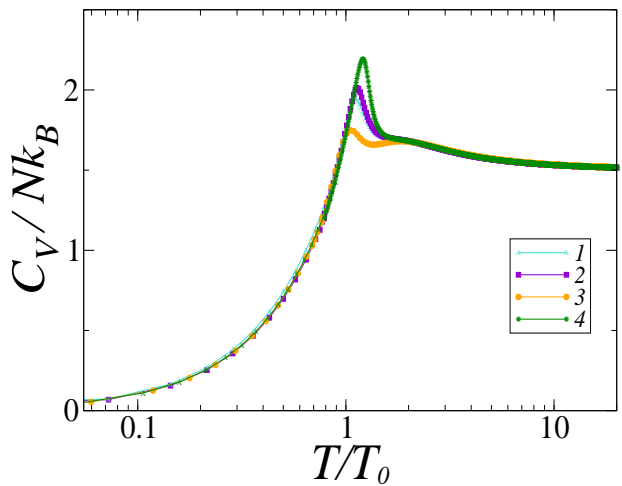}}
\caption{(Color online) Specific heat for random 90 percent variations in the strength of the barriers for a system of $M=10$ ordered planes. Each curve denotes a different set of random trials}
\label{fig:Cvcompositional}
\end{figure}

{The effects of disorder in the specific heat of the system may be analyzed by
introducing random variations in the positions of the barriers (structural
disorder) through $z_{j}\rightarrow (j+\delta_{j})a$ with $\left\vert \delta_{j}\right\vert <1,$ a random
number. Alternately, a random variation on the strengths of the barriers can
mimic compositional disorder by setting $\Lambda
_{j}\rightarrow (1+\delta_{j})\Lambda $. In the former case, we probe a sample of $M=10$ barriers at varying positions. The specific heat for
several trials in the choice of $\delta_{j}$ for the positions of the layers is shown in
Fig. \ref{fig:Cvstructural} where a 10 percent random variation is adopted. The case of compositional disorder is shown in Fig
\ref{fig:Cvcompositional} for a 90 percent random variation of $\delta_{j}$ in the strengths of the delta potentials. In
both cases, the maximum of the specific heat increases and its temperature is always higher than the corresponding values of the reference, ordered system, as well as the \textit{average} curve from several random trials labeled 1,2,3,4 in the figures. However, when
vacancies are introduced a dramatic effect in the specific heat is observed
as shown in the next Section. }
\vspace{1.0cm}
%
%
%
%
%
%
\section{Effect of vacancies}

\begin{figure}[tbh]
\centerline{%
\epsfig{file=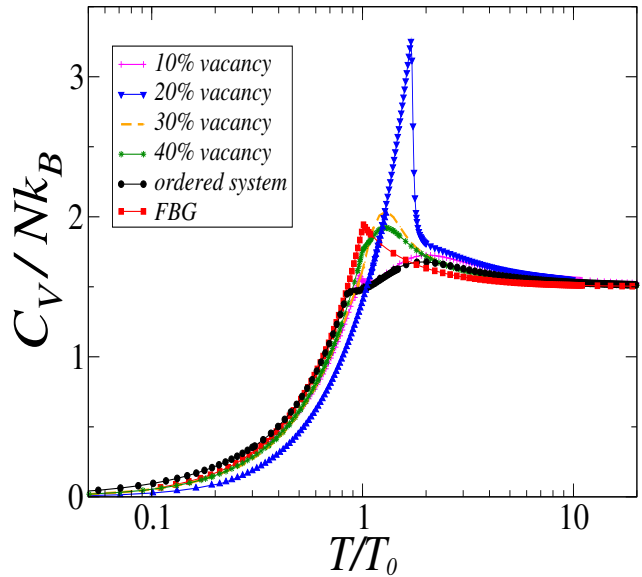,height=3.30in,width=3.0in,angle=-90}}
\caption{(Color online) $M=100$, $P_0=10$, $a/\protect\lambda_0=1$. Specific heat for a
system with 100 delta-function planes (ordered system) and with a proportion
of removed planes from symmetric sites.}
\label{fig:n100sim}
\end{figure}

\begin{figure}[bth]
\vspace{0.50cm}
\centerline{
\epsfig{file=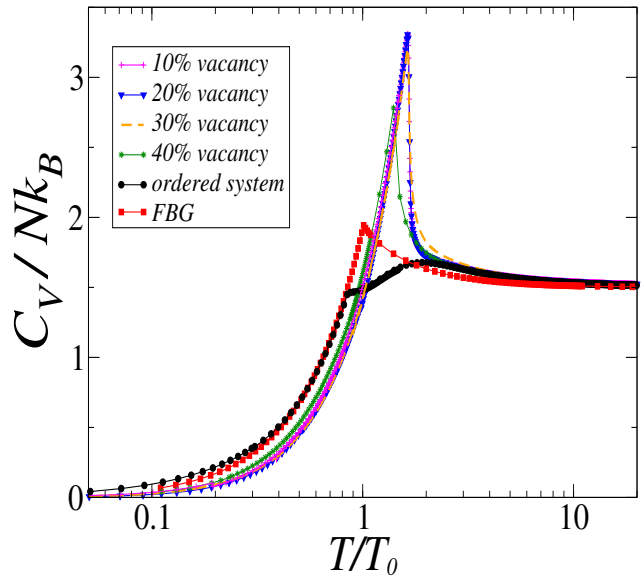,height=3.30in,width=3.in,angle=-90}}
\caption{(Color online) $M=100$, $P_0=10$, $a/\protect\lambda_0=1$. Same as the preceding
figure but with a random proportion of removed planes.}
\label{fig:n100des}
\end{figure}

{\ Using a small sample of $M=10$ planes, we calculate $C_{V}$ when a number
of vacancies is produced by setting $\Lambda _{j}=0$ for some $j^{\prime}$s. 
We start by removing walls in progressive positions. 
 When one wall is removed, a level in each energy package of the ordered case jumps down thereby modifying the energy gap between the lowest level and the following one. This phenomenon produces an effective energy gap in the first set of energy levels whose magnitude depends on the position of the removed wall. 

\begin{table}[tbh]
\begin{tabular}{|c|c|c|c|}
\hline
$T_{max}$ \quad & $C_V(T_{max})$ \quad & gap [$\hbar^2/2ma^2$] & site \\
\hline
1.39100100 & 2.590968641 & 4.617840120 & 1 \\ \hline
1.40013969 & 2.610722976 & 4.741202730 & 2 \\ \hline
1.40019796 & 2.610862299 & 4.762015852 & 3 \\ \hline
1.40019848 & 2.610863260 & 4.791629539 & 4 \\ \hline
1.40019849 & 2.610863267 & 4.835378530 & 5 \\ \hline
\end{tabular}%
\caption{Specific heat height $C_V(T_{max})$ at $T_{max}$ for systems with one site vacancy indicated in column 4.}
\end{table}
The largest gap appears when one layer near the center of the sample is removed. In Table 1 we show the specific heat maxima $C_V(T_{max})$ as well as the temperature $T_{max}$ where the specific heat attains its maximum, and the magnitude of the energy gap when the layer in a specific site is removed. The gap, $C_V(T_{max})$ and $T_{max}$ increases as the position of the removed layer varies from site 1 (near the edge) to site 5.
We claim that the increase in $C_V(T_{max})$ and $T_{max}$ is caused by the appearance of this  gap since for a 3D infinite ideal Bose gas with a quadratic dispersion relation plus a gap, both the BEC critical temperature and specific heat height at its critical temperature increase as a function of the gap magnitude \cite{MA}. Since our system is semi-infinite 
the specific heat is unable to develop a sharp peak which is a signature of a phase transition. Instead a pronounced maximum is observed which suggests a precursor of a BEC transition since it becomes sharper as the size of the system grows as shown in Figs. \ref{fig:n10several} to \ref{fig:n100des}. 
The removal of additional walls has a comparable effect in the energy levels. Each suppression moves down one level and if the removed walls were in symmetrical sites, the lowest levels are degenerate.  
\begin{figure}[tbh]
\centerline{%
\epsfig{file=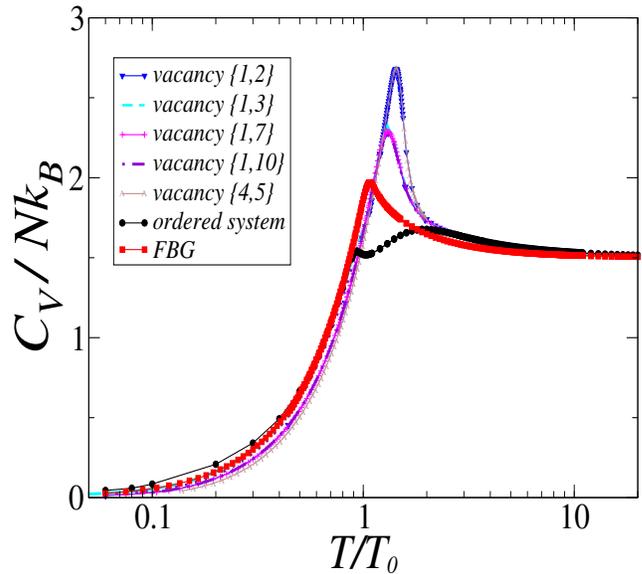,height=3.30in,width=3.0in,angle=-90}}
\caption{(Color online) Specific heat for a
system of 10 delta-function planes (ordered system) and with the indicated
vacancies with $\Lambda=10$, $a/\protect\lambda_0=1$. The curve labeled $FBG$ is the specific heat of a free boson gas
confined in the finite box.}
\label{fig:n10pairs}
\end{figure}
\begin{figure}[tbh]
\epsfig{file=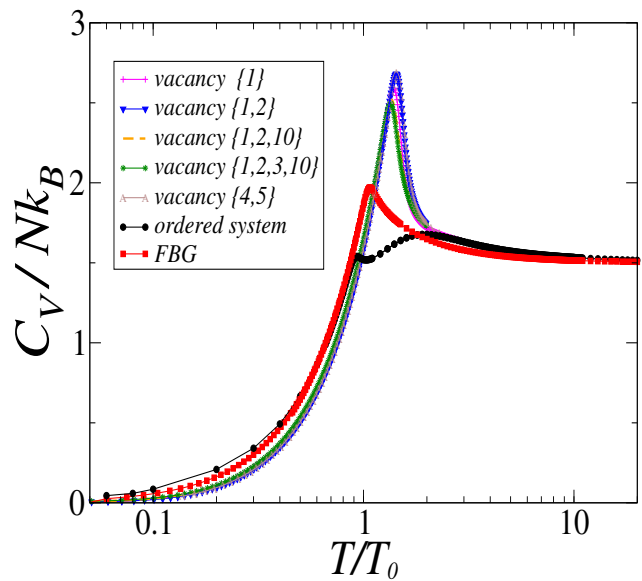,height=3.30in,width=3.0in,angle=-90}
\caption{(Color online) Specific heat for a
system with 10 delta-function planes with indicated vacancies with $\Lambda=10$, $a/\protect\lambda_0=1$.}
\label{fig:n10several}
\end{figure}

A similar behavior is observed in a larger sample with $M=100$ planes as
shown in Figs. \ref{fig:n100sim} and \ref{fig:n100des}.
In Fig. \ref{fig:n10pairs} we show the effects when there are two
vacancies compared to the ordered case and the free boson gas. The removal
at the edge of one or two planes shows a dramatic increase in the first
maximum of the specific heat at somewhat larger temperatures. This maximum is even more pronounced when there are two consecutive vacancies in the middle of the sample.
Figure \ref{fig:n10several} shows the effects of additional vacancies in the system.
}
The specific heat in the ordered system has
two maxima as a function of temperature. The maximum at the lower temperature is a signature of a BEC transition for a Bose gas confined in an infinite layered structure that would be present in the infinite system which appears at a critical temperature lower than $T_0$ as discussed in Ref. \cite%
{SalasPRA2010}. The second
maximum signals the onset of the BEC of an ideal Bose gas between two consecutive walls. 
When vacancies are introduced, either at symmetric or at random sites as shown in Figs. \ref{fig:n100sim} and \ref{fig:n100des}, respectively, the second maximum is suppressed and the first peak has a sharp increase both, in its magnitude and in the temperature.

\section{Conclusions}

Lattice disorder is known to produce localization in Fermi and, more recently, in Bose  systems in the thermodynamic limit. In this work, we have studied the effects of lattice disorder in the specific heat of finite, Bose systems by introducing random variations in
the positions and/or strengths of the planes. The overall effect of these
variations is to
modify the position, as a function of the temperature, of the maximum of the specific heat to higher temperatures. Its magnitude also increases when a large number of random samples is considered. However, when disorder is caused by the suppression of a random number of planes in the $z-$direction a dramatic effect is observed. In the case of a small sample with $M=10$ walls, the lower temperature maximum of the specific heat
increases by a large amount and it occurs at larger temperatures compared to the ordered and the free Bose gas cases. In addition, the maximum at
higher temperatures disappears. As more walls are removed, 
the magnitude of the first maximum decreases.

The case of a larger sample ($M=100$) reveals a similar behavior albeit in a
more dramatic way. Vacancies were introduced in symmetric 
or random 
configurations. In both
cases, the removal of about 20 walls increased the magnitude of the maximum
of $C_{V}$ by twice the value in the ordered system and it appeared at
higher temperatures. This behavior indicates that the large increase in the magnitude of
the specific heat maximum is caused by the appearance of a gap in the
energy spectrum. It also suggests 
that a critical number of plane vacancies promotes the emergence of a
condensate phase at a temperature above the IBG critical temperature.

This work was partially supported by grants UNAM-DGAPA IN-111613 and CONACyT 221030.


\begin{thebibliography}{99}

\bibitem{SalasPRA2010} {P. Salas, et al., Phys. Rev. A, \textbf{82}, 033632 (2010).}

\bibitem{SalasJLTP2010} {P. Salas, et al., J. Low Temp Phys \textbf{159}, 540 (2010).}

\bibitem{Pathriax} H. R. Pajkowski and R. K. Pathria, J. Phys. A: Math. Gen. 10 561 (1977); R. K. Pathria. Phys. Rev. A 5, 1451 (1972).

\bibitem{Otros} D. M. Goble and L. E. H. Trainor, Phys. Rev. 157, 167 (1967); M. F. M. Osborne, Phys. Rev. 76, 3965 (1949).

\bibitem{Chakravarty} {S. Chakravarty, A. Sudb$\emptyset$, P. W. Anderson, S. Strong, Science, \textbf{261}, 337 (1993).}

\bibitem{surfacestates} {W.L. Bloss, Phys. Rev. B \textbf{44}, 8035 (1991); W. Shockley, Phys. Rev. {\bf 56}, 317 (1939); I. Tamm, Physik Zeits Sowjetunion {\bf 1}, 733 (1932)}.

\bibitem{Griffiths} {D. J. Griffiths and C.A. Steinke, Am. J. Phys. \textbf{69}, 137 (2001).}

\bibitem{Anderson} {Philip Anderson, Absence of Diffusion in Certain Random
Lattices Phys. Rev. \textbf{109}, 1492 (1958).}

\bibitem{Roati} {Giacomo Roati, Chiara D'Errico, Leonardo Fallani, Marco
Fattori, Chiara Fort, Matteo Zaccanti, Giovanni Modugno, Michele Modugno, and Massimo Inguscio, Letter to Nature \textbf{453}, 895  (2008).}

\bibitem{Modugno} Michele Modugno, New Journal of Physics, \textbf{11}, 033023 (2009).

\bibitem{Aspect} {Juliette Billy, Vincent Josse, Zhanchun Zuo, Alain Bernard, Ben Hambrecht, Pierre Lugan, David Cl\'{e}ment, Laurent Sanchez-Palencia, Philippe Bouyer, and Alain Aspect, 
Nature \textbf{453}, 891 (2008).}

\bibitem{Aspect2} {L. Sanchez-Palencia, D. Cl\'{e}ment, P. Lugan, P. Bouyer, G. V. Shlyapnikov, and A. Aspect, Anderson, 
Phys. Rev. Lett. \textbf{98}, 210401 (2007).}

\bibitem{Luttinger1} {J.M. Luttinger and H.K. Sy, 
Phys. Rev. A, \textbf{7},701 (1973).}

\bibitem{Luttinger2} {J.M. Luttinger and H.K. Sy, 
Phys. Rev. A, \textbf{7},712 (1973).}

\bibitem{H-M-W} {P.C. Hohenberg, Phys. Rev. \textbf{158}, 383, (1967); N.D. Mermin and H. Wagner, Phys Rev. Lett. \textbf{17}, 1133  (1966)}

\bibitem{MA} M.A. Sol\'is, ``Tamm's surface states and Bose-Einstein condensation", 
in preparation.

\bibitem{Pathria} R. K. Pathria, Statistical Mechanics, 2nd ed. (Pergamon, Oxford, 1996) p506.

\end{thebibliography}
\end{document}